\documentclass{ifacconf}
\usepackage{natbib} 

\usepackage{amsmath,amssymb,amsfonts}
\usepackage{mathtools}
\usepackage{nicefrac}
\usepackage{dsfont}
\usepackage{algorithmic}
\usepackage{graphicx}
\usepackage{textcomp}
\usepackage{xcolor}
\usepackage[nolist,nohyperlinks]{acronym}
\def\BibTeX{{\rm B\kern-.05em{\sc i\kern-.025em b}\kern-.08em
    T\kern-.1667em\lower.7ex\hbox{E}\kern-.125emX}}

\usepackage{booktabs}
\usepackage{makecell}
\usepackage{csquotes}
\usepackage{url}
\usepackage{siunitx} 

\definecolor{vgRed}{RGB}{193, 48, 24}
\definecolor{vgOrange}{RGB}{243, 111, 19}
\definecolor{vgYellow}{RGB}{235, 203, 56}
\definecolor{vgGreen}{RGB}{162, 185, 105}
\definecolor{vgLightBlue}{RGB}{13, 149, 188}
\definecolor{vgDarkBlue}{RGB}{6, 56, 81}

\definecolor{hks}{RGB}{199,16,92}
\definecolor{Dunkelgr}{RGB}{21, 56, 36}
\definecolor{Hellblau}{RGB}{80, 149, 200}
\definecolor{Gelbgrue}{RGB}{196, 210, 15}
\definecolor{Hellgrue}{RGB}{74, 172, 150}
\definecolor{Goldgelb}{RGB}{234, 195, 114}

\definecolor{cbBlue}{HTML}{0072B2}
\definecolor{cbOrange}{HTML}{E69F00}
\definecolor{cbGreen}{HTML}{009E73}

\newtheorem{remark}{Remark}

\newcommand{\imag}{\ensuremath{\mathrm{\imath}}}

\begin{document}
\begin{acronym}
    \acro{ml}[ML]{machine learning}
    \acro{nn}[NN]{neural network}
    \acro{gnn}[GNN]{graph neural network}
    \acro{mlp}[MLP]{multilayer perceptron}
    \acro{lstm}[LSTM]{long short-term memory}
    \acro{tcn}[TCN]{temporal convolutional network}
    \acro{stgcn}[STGCN]{spatio-temporal graph convolutional network}

    \acro{ho}[HO]{hyperparameter optimization}
    \acro{ivp}[IVP]{initial value problem}
    \acro{ode}[ODE]{ordinary differential equation}
    \acro{node}[NODE]{neural ordinary differential equation}
    \acro{dnode}[D-NODE]{data-controlled neural ordinary differential equation}
    \acro{anode}[A-NODE]{augmented neural ordinary differential equation}
    \acro{mpgnode}[MPG-NODE]{message passing graph neural ordinary differential equation}

    \acro{pinn}[PINN]{physics informed neural network}
    \acro{nhnn}[NHNN]{nearly hamiltonian neural network}
    \acro{sindy}[SINDy]{sparse identification of nonlinear dynamics}
    \acro{dp5}[DOPRI5]{fifth-order Dormand-Prince}
    \acro{rk4}[RK4]{fourth-order Runge-Kutta}
    \acro{der}[DER]{distributed energy resource}
    \acro{sg}[SG]{synchronous generator}
    \acro{gfi}[GFI]{grid-forming inverter}
    \acro{smib}[SMIB]{single-machine-infinite-bus}
    \acro{nhnn}[NHNN]{nearly hamiltonian neural network}
    \acro{rmse}[RMSE]{root-mean-square-error}
    \acro{mse}[MSE]{mean squared error}
    \acro{relu}[ReLU]{rectified linear unit}
    \acro{gelu}[GELU]{Gaussian error linear unit}
    \acro{silu}[SiLU]{sigmoid linear unit}

    \acro{io}[I/O]{input/output}

    \acro{ieee9}[IEEE9]{IEEE 9 bus system}
    \acro{ieee30}[IEEE30]{IEEE 30 bus system}
    \acro{ieee39}[IEEE39]{IEEE 39 bus system}
\end{acronym}

\begin{frontmatter}
\title{Graph Neural Ordinary Differential Equations for Power System Identification with Transfer Learning Applications} 

  
\author[First]{Hannes M. H. Wolf}
\author[First]{Christian A. Hans} 
  
\address[First]{Automation and Sensorics in Networked Systems Group, University of Kassel, Germany (e-mail: \{h.wolf, hans\}@uni-kassel.de)}


\begin{abstract}  
    With the shift towards decentralized energy generation, the increasing complexity of power systems renders physics-based modeling challenging. 
    At the same time the growing amount of available measurement data opens the door for obtaining data-driven models.  
    A modern method to do so are \acp{node}, offering a framework for continuous-time system identification. 
    Recent extensions, so called graph \acp{node} impose a structural inductive bias that has the potential to improve generalization of the learned representations. 
    In this work, we employ and extend them to develop message-passing graph \acp{node} (\acsp{mpgnode}) for identification of coupled systems with heterogeneous node dynamics and heterogeneous edge couplings. 
    This includes state-of-the-art \acl{ml} architectures to infer latent representations of unmeasured states, local node and edge embeddings to account for heterogeneity as well as an autoregressive scheme to allow for piecewise constant control inputs. 
    We apply \acsp{mpgnode} to identify voltage and frequency dynamics of power systems and compare them to a monolithic \ac{node} under identical measurement assumptions. 
    Our case study indicates that while achieving similar performance, our proposed \acsp{mpgnode} are more flexible with transfer learning options that allow to add or remove edges and nodes with little retraining. 
\end{abstract}

\end{frontmatter}

\acresetall

\section{Introduction}
\label{sec:introduction}
In power systems, dynamical models of frequency and voltage transient behaviour are essential for stability analysis, control design and prediction. 
Traditionally such models are derived from first principles. 
However, due to an increasing number of distributed generation units, systems become more complex, which complicates physics-based modeling. 
At the same time, the growing availability of measurement data enables the use of data-driven system identification approaches. 

\Acp{gnn} and \acp{node} are two data-driven approaches that align well with power systems.
\Acp{node} provide a natural framework for identifying continuous-time systems. 
In recent works \citep{aryal_neuralps_2023, wolf_ident_2025, zhang_learning_2024}, they have been applied to identify power system dynamics. 
However, aforementioned studies require synchronized voltage phase angle measurements, which can be hard to obtain in practice. 
\cite{wolf_augmented_2025} illustrate how to drop this requirement by inferring latent phase angle representations from a sequence of recent observations. 
\Acp{gnn} can leverage power systems' inherent network structure. 
Recent works \citep{nandanoori_graph_2022,zhao_structure-informed_2022,yu_pidgeun_2024,wang_multi-task_2025} have employed spatio-temporal convolutional \acp{gnn} as well as message-passing \ac{gnn} combined with attention mechanisms to predict state trajectories of power systems using past observations. 
However, these models are discrete-time forecasting architectures that lack control inputs, making them hard to use for analysis and control design.

In this work we combine the continuous-time formulation of \acp{node} and the structural inductive bias from \acp{gnn} to propose message-passing graph \acp{node} (\acsp{mpgnode}) for system identification. 
In general, the idea is to represent node dynamics and edge couplings by neural networks, that are shared across all nodes and edges in a known graph topology. 
This has the potential to improve generalization, thereby facilitating training, and to provide a flexible model for transfer learning under topology changes. 
Methodologically, vanilla graph \acp{node} \citep{poli_graph_2019} are the basis for our approach.
Motivated by \cite{allier_decomposing_2025} and \cite{huang_generalizing_2023} we extend them to consider heterogeneous node dynamics in neural graph dynamical models. 
More specifically, \cite{huang_generalizing_2023} propose graph \acp{node} together with graph encoders to obtain environment embeddings that condition shared node dynamics, and consider fully latent initial states from global measurements. 
\cite{allier_decomposing_2025} propose per-node embeddings, that act as additional inputs to the \ac{nn} and condition the node dynamics while assuming full state measurements. 
In our approach, we adopt the idea of per node embeddings from \cite{allier_decomposing_2025}. 
However, we use continuous-time formulations that employ \acp{node} instead of state derivative predictors. 
Furthermore, our approach does not require full state measurements and instead infers unmeasured states from past measurements via \acp{stgcn}. 
Also, we preserve measured states of the original system in the state-space of the \ac{node}, which helps to regularize the model in a physics-inspired manner. 
As opposed to \cite{huang_generalizing_2023}, we use persistent node embeddings instead of sample-specific embeddings obtained from past measurements to account for heterogeneity among nodes with time-invariant local dynamics. 

Due to their application domain, both approaches \cite{allier_decomposing_2025} and \cite{huang_generalizing_2023} do not consider heterogeneous edge dynamics or control inputs. 
As opposed to that, we equip our model with learnable edge embeddings that condition the message-passing function, as well as control inputs, that drive the node dynamics. 
Overall this allows us to move more closely to general nonlinear state model descriptions that are widely considered in the control community.

Our contributions are as follows:
1) We develop control-oriented \acp{mpgnode} for the identification of coupled systems with local dynamics that allow for heterogeneous node dynamics and edge couplings through learned node and edge embeddings. 
Our model uses an \ac{stgcn} to infer partially latent initial states from past observations.
Furthermore, we account for discontinuous control inputs by unrolling the model in an autoregressive fashion.  
2) We apply our approach to identify the dynamics of power systems, leveraging their given graph structure as an inductive bias for identification. 
Specifically, we infer latent phase-angle representations from historic observations, thereby avoiding the need for synchronized phasor measurements and use learned embeddings to account for hetergeneous unit dynamics and line couplings. 
3) We compare the proposed \ac{mpgnode} to a monolithic \ac{node} baseline, as well as a homogeneous \ac{mpgnode} baseline on an altered IEEE 9-bus system. 
Moreover, we demonstrate the transfer learning capabilities of the \ac{mpgnode} under topology changes, i.e., adding and removing nodes and edges.

The paper is structured as follows. 
Section \ref{sec:preliminaries} introduces basic notation and fundamental neural architectures. 
We then proceed with the system modeling in Section \ref{sec:modeling} and the proposed \ac{mpgnode} in Section \ref{sec:nodes_for_system_identification}. 
Section \ref{sec:case_study} presents the case study. Section \ref{sec:conclusion} concludes the paper.

\section{Preliminaries}
\label{sec:preliminaries}
\subsection{Notation}
\label{sec:notation}

We denote the set of real numbers by $\mathbb{R}$, the set of non-negative real numbers by $\mathbb{R}_{\geq 0}$ and the set of strictly positive real numbers by $\mathbb{R}_{> 0}$. 
The set of positive integers is denoted by $\mathbb{N}$. 
An element of the set of complex numbers $\mathbb{C}$ is denoted by $z = a + \imag b$ with $a, b \in \mathbb{R}$ and where $\imag$ is the imaginary unit. 
The cardinality of a finite countable set $\mathcal{S}$ is $|\mathcal{S}|$.
For phasors, we make use of polar notation, i.e., $z=|z|e^{\imag \varphi}$ where $|z| \in \mathbb{R}$ is the magnitude and $\varphi \in \mathbb{R}$ is the phase angle. 
Datasets used in supervised learning tasks are denoted by $\mathcal{D} = \{(\xi_n, \eta_n)\}_{n=1}^N = \{(\xi_1, \eta_1), \ldots, (\xi_N, \eta_N)\}$, $N \in \mathbb{N}$. 
One element of such a dataset is called sample, where $\xi_n$ are the input features, i.e., the input to the \ac{ml} model and $\eta_n$ is the target, i.e., the desired output of the \ac{ml} model. 
Sampling $c$ from a uniform distribution in an interval $[a, b]$ is denoted by $c \sim \mathcal{U}(a, b)$.

\subsection{Graphs}
\label{sec:graphs}

A graph $\mathcal{G} = (\mathcal{V}, \mathcal{E})$ consists of a set of $|\mathcal{V}|$ nodes $i \in \mathcal{V}$ and a set of edges $(i,j) \in \mathcal{E} \subset \mathcal{V} \times \mathcal{V} \setminus \{(i,i)\}_{i=1}^{|\mathcal{V}|}$. 
The set of neighbors of node $i$ is $\mathcal{N}_i = \{j \in \mathcal{V} | (i,j) \in \mathcal{E}\}$. 
The adjacency matrix $A \in \mathbb{R}^{N \times N}$ of a graph $\mathcal{G}$ is defined as $A_{ij} = 1$ if $(i,j) \in \mathcal{E}$ and $A_{ij} = 0$ otherwise. 
Artificially adding self-loops to each node yields $\tilde{A} = A + I_N$.
Furthermore, $S = D^{\nicefrac{-1}{2}} \tilde{A} D^{\nicefrac{-1}{2}}$ is the normalized adjacency matrix where $D$ is the degree matrix with $D_{ii} = (\sum_{j=1}^N \tilde{A}_{ij})$ and $D_{ij} = 0$ for $i \neq j$.

\subsection{\Aclp{nn}}
\label{sec:neural_networks}

\subsubsection{\Aclp{mlp}:}
\label{sec:mlp}
\Acp{mlp} are fully connected feed-forward \acp{nn} composed of one input layer, $L \in \mathbb{N}$ hidden layers and one output layer. 
The mapping between consecutive layers is given by an affine transformation, which is followed by an activation function $\sigma$ in the hidden layers. 
The input $\xi$ to such a network is mapped to the output via 
\begin{equation}
    \label{eq:mlp}
    f_{\theta}(\xi) = W^{L+1}\sigma(W^{L} \sigma (\dots \sigma(W^1 \xi + b^1) \dots )+ b^{L}) + b^{L+1}, 
\end{equation}
where $W^l \in \mathbb{R}^{n_{l-1} \times n_l}, n_l \in \mathbb{N}$ are the weights and $b^l \in \mathbb{R}^{n_l}$ are the biases of layer $l \in [1, L+1] \subset \mathbb{N}$. 
Together, they form the parameters $\theta$ of the \ac{mlp}. 

\subsubsection{\Aclp{stgcn}:}
\label{sec:tcn}
\Acp{stgcn} \cite{yu_stgcn_2018} capture temporal and spatial dependencies in sequential data over a graph structure. 
In what follows, we introduce \acp{stgcn} starting from 1D-causal temporal convolutional layers, moving over to graph convolutional layers, the construction of residual blocks and the overall architecture.
For this, consider two multi-node, multi-channel sequences $\xi_{\text{in}} = \{\xi_{\text{in}}(t_1), \allowbreak \dots, \xi_{\text{in}}(t_T)\}$ and $\xi_{\text{out}} = \{\xi_{\text{out}}(t_1),\allowbreak \dots, \xi_{\text{out}}(t_T)\}$, where $T$ is the length of the sequences, $\xi_{\text{in}}(t_k) \in \mathbb{R}^{|\mathcal{V}| \times C_{\text{in}}}$ the input and $\xi_{\text{out}}(t_k) \in \mathbb{R}^{|\mathcal{V}| \times C_{\text{out}}}$ the output at time $t_k$. 
In what follows, $\xi_{\text{in}}^{i, c}(t_k) \in \mathbb{R}$ denotes the input at channel $c \in [1, C_{\text{in}}] \subset \mathbb{N}$ and node $i \in [1, |\mathcal{V}|] \subset \mathbb{N}$, whereas $\xi_{\text{out}}^{i, r}(t_k) \in \mathbb{R}$ denotes the output at channel $r \in [1, C_{\text{out}}] \subset \mathbb{N}$ and node $i$. 

Temporal dependencies are captured by 1D-causal convolutional layers. 
Consider an input sequence $\xi_{\text{in}}$. 
The output at channel $r$, node $i$ and time $t_k$ is then given by
\begin{equation}
    \label{eq:tcn_conv}
    \xi_{\text{out}}^{i, r}(t_k) = \hat{b}^{r} + \sum_{c=1}^{C_{\text{in}}} \sum_{q=0}^{Q-1}  \hat{W}_{c,q}^{r} \, \xi_{\text{in}}^{i, c}(t_{k - d  q})
\end{equation}
where $d \in \mathbb{N}$ is the dilation, i.e., a temporal spacing between elements in the convolution. Moreover, weights $\hat{W}^{r} \in \mathbb{R}^{C_{\text{in}} \times Q}$, where $Q$ is the kernel size, and bias $\hat{b}^{r} \in \mathbb{R}$ are parameters of the temporal convolutional layer associated with channel $r$ which are shared across all nodes. 

Spatial dependencies are captured by graph convolutional layers.
At each time instant of an input sequence $\xi_{\text{in}}(t_k)$, the output of the graph convolutional layer is 
\begin{equation}
    \label{gcn_conv}
    \xi_{\text{out}}(t_k) = S \, \xi_{\text{in}}(t_k) \, \tilde{W}  +  \mathds{1} \, \tilde{b}^T 
\end{equation}

Moreover, weights $\tilde{W} \in \mathbb{R}^{C_{\text{in}} \times C_{\text{out}}}$ and bias $\tilde{b} \in \mathbb{R}^{C_{\text{out}}}$ are parameters of the graph convolutional layer which are shared across all time instants.

Following standard practice, we use residual blocks that map input sequences to output sequences with
\begin{equation}
    \xi_{\text{out}} = p_b(\xi_{\text{in}}) + h_b(\xi_{\text{in}}),
\end{equation}
where $h_b$ of block $b \in [1, B] \subset \mathbb{N}$ consists of two temporal convolutional layers \eqref{eq:tcn_conv} with one graph convolutional layer \eqref{gcn_conv} between them and each layer is followed by a rectified linear unit. 
Furthermore, $p_b$ is an affine projection layer accounting for potential dimension mismatches.

We consider \Acp{stgcn} that consist of $B \in \mathbb{N}$ consecutive residual blocks where temporal convolutional layers' dilation grows exponentially, i.e.,  $d_b = 2^{b-1}$. 
In the presented form, the temporal receptive field of \acp{stgcn} is then given by $R = 1 + 2 (Q - 1) (2^{B} - 1)$. 
Furthermore, one block $b$ extends the spatial receptive field by one hop.

\section{Modeling}
\label{sec:modeling}
We model a power system as a graph and associate each node $i \in \mathcal{V}$ with the following quantities: 
A unit injecting or consuming active power $p_i \in \mathbb{R}$ and reactive power $q_i \in \mathbb{R}$, a voltage phasor $v_i e^{\imag \varphi} \in \mathbb{C}$ with magnitude $v_i \in \mathbb{R}_{\geq0}$ and angle $\varphi_i \in \mathbb{R}_{\geq 0}$, rotating at the angular frequency $\omega_i \in \mathbb{R}_{\geq 0}$, as well as a local rotating $dq$-coordinate-system with angle $\varphi_i\in \mathbb{R}_{\geq 0}$ whose $q$-axis is aligned with the voltage phasor. 
The relative angle of this coordinate system with respect to a global reference frame with angle $\varphi_{\text{ref}}$ is given by $\delta_i = \varphi_i - \varphi_{\text{ref}}$ and the relative angular velocity by $\dot{\delta}_i = \omega_i - \omega_{\text{ref}}$ \citep{schiffer_stability_2015}. 
Lastly, we associate a parallel admittance $y_{ii} = g_{ii} + \imag b_{ii} \in \mathbb{C}$ with conductance $g_{ii} \in \mathbb{R}_{\geq 0}$ and susceptance $b_{ii} \in \mathbb{R}$ to each node. 
Admittances $y_{ii}$ comprise all shunt elements and constant impedance loads connected to node $i$. 
Each edge $(i,j) \in \mathcal{E}$ is associated with a series admittance $y_{ij} = g_{ij} + \imag b_{ij} \in \mathbb{C}$, $g_{ii} \in \mathbb{R}_{\geq 0}$ and $b_{ii} \in \mathbb{R}$. 

\subsection{Power flow model}
Consider nodes $i$ and $j$ connected by edge $(i,j) \in \mathcal{E}$. 
The active and reactive power flow from $i$ to $j$ is given by \citep{kundur_power_1994}
\begin{align}
    p_{ij} &= g_{ij} v_i^2 - g_{ij} v_i v_j \cos(\delta_{ij}) - b_{ij} v_i v_j \sin(\delta_{ij}), \\
    q_{ij} &= -b_{ij} v_i^2 + b_{ij} v_i v_j \cos(\delta_{ij}) - g_{ij} v_i v_j \sin(\delta_{ij}),
\end{align}
where $\delta_{ij} = \delta_i - \delta_j$ is the angle difference between the two nodes. 
Furthermore, the consumed active and reactive power of all load and shunt elements at node $i$ is $p_{ii} = v_i^2  g_{ii}$ and $q_{ii} = -v_i^2  b_{ii}$, respectively. 
Collecting the power flows of node $i$ to all neighbors, as well as the power consumed by all load and shunt elements, we obtain the injected active and reactive power at node $i$,
\begin{equation} 
\label{eq:power_injections} 
\begin{aligned}
    \hspace{-0.3em} p_{i}\hspace{-0.1em} &= \hspace{-0.1em} \, v_i^2 \, g_{ii} \hspace{-0.1em} + \hspace{-0.1em} v_i \hspace{-0.4em} \textstyle \sum\limits_{j \in \mathcal{N}_i} \hspace{-0.4em} v_i g_{ij} \hspace{-0.1em} - \hspace{-0.1em} v_j\left(g_{ij} \cos (\delta_{ij}) \hspace{-0.1em} + \hspace{-0.1em}b_{ij} \sin (\delta_{ij})\right),\\
    \hspace{-0.3em} q_{i} \hspace{-0.1em}&= \hspace{-0.1em}- v_i^2 \, b_{ii} \hspace{-0.1em} - \hspace{-0.1em} v_i \hspace{-0.4em} \textstyle\sum\limits_{j \in \mathcal{N}_i}\hspace{-0.4em} v_i b_{ij} \hspace{-0.1em} + \hspace{-0.1em}v_j\left(g_{ij} \sin (\delta_{ij}) \hspace{-0.1em} - \hspace{-0.1em}b_{ij} \cos (\delta_{ij})\right). \mkern-50mu
\end{aligned}
\end{equation}

\subsection{Unit model}
We consider droop-controlled \acp{gfi} at the nodes. 
The frequency and voltage of each unit are adjusted according to active and reactive power deviations. 
In detail, we use droop-control laws of the form \cite{guerrero_advanced_2013}
\begin{subequations}
\label{eq:droop_laws}
\begin{align}
    \label{eq:frequency_droop_law}
    \omega_i&= \omega_i^d -k_i^p p^\Delta_i, \\
    \label{eq:voltage_droop_law}
    v_i  &= v_i^d - k_i^q q^\Delta_i,
\end{align}
\end{subequations}
where $k_i^p \in \mathbb{R}_{> 0}$ and $k_i^q \in \mathbb{R}_{> 0}$ are the droop gains and $w_i^d \in \mathbb{R}_{> 0}$
and $v_i^d \in \mathbb{R}_{> 0}$ are the frequency and voltage setpoints, respectively. 
Furthermore, $p^\Delta_i \in \mathbb{R}$ and $q^\Delta_i \in \mathbb{R}$ are the low-pass filtered power deviations between active power setpoint $p_i^d \in \mathbb{R}$ and active power $p_i \in \mathbb{R}$, as well as between reactive power setpoint $q_i^d \in \mathbb{R}$ and reactive power $q_i \in \mathbb{R}$.
Following standard practice \citep{schiffer_conditions_2014}, a first order low pass filter with time constant $\tau_i \in \mathbb{R}_{> 0}$ is used, i.e., 
\begin{subequations}
\label{eq:filter_laws}
\begin{align}
    \tau_i \dot{p}^\Delta_i &= - p^\Delta_i + (p_i - p_i^d), \\
    \tau_i \dot{q}^\Delta_i &= - q^\Delta_i + (q_i - q_i^d).
\end{align}
\end{subequations}
Combining \eqref{eq:droop_laws} and \eqref{eq:filter_laws} under the assumption that $\dot{\omega}^d_i = 0$, $\dot{v}^d_i = 0$ and considering $\dot{\delta}_i = \omega_i - \omega_{\text{ref}}$, we obtain a state model for the unit dynamics of the form \cite{schiffer_conditions_2014}
\begin{subequations}
\label{eq:unit_model}
\begin{align}
    \dot{\delta}_{i} &= \omega_i - \omega_{ref}, \\
    \dot{\omega}_i &= \frac{1}{\tau_i} \big(- \omega_i + \omega_i^d- k_i^p (p_{i} - p_i^d) \big), \label{eq:frequency_dynamics}\\
    \dot{v}_i &= \frac{1}{\tau_i} \big( -v_i + v_i^d - k_i^q (q_{i} - q_i^d )\big) .
\end{align}
\end{subequations}
\begin{remark}
    Note that choosing $\tau_i = k_i^p \, m_i$, where $m_i$ is the inertia constant of a \ac{sg}, the frequency dynamics of the inverter are identical to the frequency dynamics of a droop-controlled \ac{sg} \citep{schiffer_synchronization_2013}, when
    omitting governor dynamics. Thus, we can use the same model for both \acp{gfi} and \acp{sg}.
\end{remark}

\subsection{Overall model}
\label{sec:overall_model}
Assigning each node $i \in \mathcal{V}$ the dynamics \eqref{eq:unit_model} and replacing $p_i$ and $q_i$ by \eqref{eq:power_injections}, we obtain a nonlinear state model of the entire system of the form
\begin{subequations}
    \label{eq:nonlinear_state_model}
\begin{align}
    \dot{x}(t) &= f(t,x(t),u(t)), \\
    y(t) &= g(x(t)),
\end{align}
\end{subequations}
where $t \in \mathbb{R}_{\geq 0}$ is the time. 
The states of the model are ${x = [\delta_1 ~ \omega_1  v_1 ~ \dots ~ \delta_{|\mathcal{V}|} ~ \omega_{|\mathcal{V}|} ~ v_{|\mathcal{V}|}]^T}$ and the inputs are ${u = [p_1^d ~ \dots ~ p_{|\mathcal{V}|}^d]^T}$. 
We assume that voltage and frequency at each node can be measured, i.e., the system outputs are  ${y = [\omega_1 ~ v_1 ~ \dots ~ \omega_{|\mathcal{V}|} ~ v_{|\mathcal{V}|}]^T}$.


\section{Graph neural ordinary differential equations}
\label{sec:nodes_for_system_identification}
\Acp{node}, first presented by \cite{chen_neural_2018}, are a class of \ac{ml} models that can be used for identification of continuous-time dynamical systems.
They are formulated as \acp{ode} in which the right hand side is modelled by a \ac{nn} $f_{\theta_f}$ with parameters $\theta_{f} \in \mathbb{R}^{n_{\theta_f}}$, i.e.,
\begin{equation}
    \label{eq:neural_ode}
    \dot{x}(t) = f_{\theta_f} (t, x(t)), \quad x(t_{s}) = x_{0}.
\end{equation}

For control-oriented system identification problems, \eqref{eq:neural_ode} can be extended by additional concepts resulting in the \acp{node} model proposed by \cite{massaroli_dissecting_2020}, i.e.,
\begin{subequations}
    \label{eq:general_neural_ode}
\begin{align}
    \dot{x}(t) &= f_{\theta_f} (t, x(t), c_{\theta_c}(\xi)), \\
    \quad x(t_{s}) &= h_{\theta_h}(\xi), \\
    y(t) &= g_{\theta_g}(x(t)).
\end{align}
\end{subequations}
Employing a numerical solver, a solution to \ac{ivp} \eqref{eq:general_neural_ode} is
\begin{align}
    \label{eq:general_solution}
    \hat{y}(t) &= g_{\theta_g}\big(h_{\theta_h}(\xi) + \int_{t_{s}}^{t_{e}} f_{\theta_f}(t, x(t), c_{\theta_c}(\xi)) dt\big).
\end{align}
\ac{ivp} \eqref{eq:general_neural_ode} extends vanilla \acp{node} \eqref{eq:neural_ode} by three concepts:
The first one is called augmentation. 
As the state of the system is usually not fully observed, we infer a (partially) latent initial state representation with $h_{\theta_h}$ from past measurements and control inputs contained in the input features $\xi \in \mathbb{R}^{n_\xi}$. 
The second concept is data control, where the dynamics are conditioned on the input features through an embedding $c_{\theta_c}(\xi)$.
From a system identification perspective, this embedding function is just a non-learnable selector of the current control input, which is assumed to be constant over $[t_s, t_e)$, from the input features $\xi$, i.e. $c_{\theta_c}(\xi)=\pi_u(\xi) = u(t_s)$. 
The last additional concept is the output function $g_{\theta_g}$ mapping the state to the output space, which follows textbook state model descriptions. 

\begin{remark}
    In this formulation, the input features $\xi$ of the system identification problem are past measurements and control inputs, as well as control inputs over $[t_s, t_e)$. 
\end{remark}
\begin{remark}
    As we consider time-invariant systems, without loss of generality we drop the time argument in $f_{\theta_f}$ in what follows. 
\end{remark}

\subsection{Message-passing graph \acp{node}}
\label{sec:graph_node}

\begin{figure*}
    \centering
    \vspace{2mm}
    \includegraphics{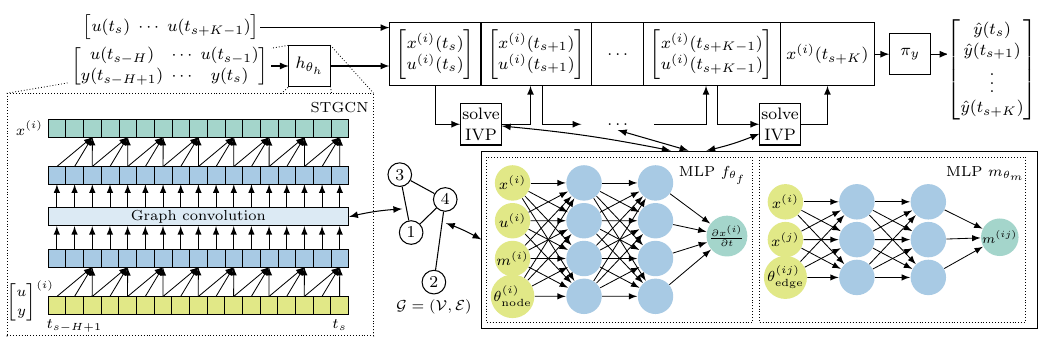}
    \caption{Schematic of the unrolled prediction model. The encoder yields an initial state for the \ac{node} from historic data. Then, multistep predictions are obtained by repeatedly solving \acp{ivp}. Note that the \acp{nn} are simplified for illustrative reasons. Typically, the state $x^{(i)}$, input $u^{(i)}$, the node embeddings $\theta^{(i)}_{\text{node}}$ and edge embeddings $\theta^{(ij)}_{\text{edge}}$ and the aggregated messages $m^{(i)}(t)$ are vectors, where each component is an input to the \ac{nn}.}
    \label{fig:neural_ode_stepper}
\end{figure*}

Many problems can be formulated as a graph $\mathcal{G} = (\mathcal{V}, \mathcal{E})$ with nodes $i \in \mathcal{V}$ and edges $(i,j) \in \mathcal{E}$. 
This inductive bias is commonly exploited in \ac{ml} by using \acp{gnn}.
\cite{poli_graph_2019} make use of static \acp{gnn} in the context of \acp{node}. 
Next to graph convolutional \acp{node}, they formulate \acp{mpgnode} $\forall i \in \mathcal{V}$ as 
\begin{align}
    \label{eq:vanilla_gmpnode}
    \dot{x}^{(i)}(t) = f_{\theta_f} (\textstyle\sum_{j \in \mathcal{N}_i} m_{\theta_m}(x^{(i)}(t), x^{(j)}(t))),
\end{align}
where $m_{\theta_m}$ is a message-passing function. 
Note that while the dynamics are computed node-wise and the message-passing edge-wise, the functions $f_{\theta_f}$ and $m_{\theta_m}$ are shared across all nodes and edges. 
In this work, we extend \eqref{eq:vanilla_gmpnode} by the concepts of \eqref{eq:general_neural_ode} as well as ideas from \cite{allier_decomposing_2025} to allow for heterogeneous node dynamics and edge couplings. 
Our \ac{mpgnode} $\forall i \in \mathcal{V}$ reads 
\begin{subequations}
\label{eq:message_passing_neural_ode}
\begin{align}
    \dot{x}^{(i)}(t) &= f_{\theta_f} (x^{(i)}(t), c_{\theta_c}(\xi^{(i)}), m^{(i)}(t),\theta_{\text{node}}^{(i)}), \\
    m^{(i)}(t) &= \textstyle\sum_{j \in \mathcal{N}_i} m_{\theta_m}(x^{(i)}(t), x^{(j)}(t), \theta_{\text{edge}}^{(ij)}), \\
    x^{(i)}(t_{s}) &= \big[h_{\theta_h}(\xi)\big]^{(i)}, \\
    y^{(i)}(t) &= g_{\theta_g}(x^{(i)}(t)).
\end{align}
\end{subequations}
Here, $m^{(i)}(t) \in \mathbb{R}^{n_m}$ is the aggregated message for node $i$ and $\theta_{\text{node}}^{(i)} \in \mathbb{R}^{n_{\theta_{n}}}$ and $\theta_{\text{edge}}^{(ij)}\in \mathbb{R}^{n_{\theta_{e}}}$ are node and edge specific learnable parameters.   
As opposed to \eqref{eq:general_neural_ode}, we apply $f_{\theta_f}$, $g_{\theta_g}$ and $c_{\theta_c}$ node-wise, with interactions entering through $m^{(i)}(t)$.
The encoder $h_{\theta_h}$ jointly processes input features $\xi$ from all nodes, where $[\cdot]^{(i)}$ selects the output for node $i$. 
Thus, we have $c_{\theta_c}(\xi^{(i)}) = \pi_u(\xi^{(i)}) = u^{(i)}(t_s)$, i.e., a non-learnable selector of the current control input at node $i$.
As opposed to \eqref{eq:vanilla_gmpnode}, we further include the state of the node itself as an argument in $f_{\theta_f}$ to allow for local dynamics for each node. 
In order to allow for heterogeneous node dynamics, we use a set of learnable parameters, called node embeddings $\theta_{\text{node}}^{(i)}$, that condition the dynamics function $f_{\theta_f}$. 
Similarly, we allow for heterogeneity of the edge couplings using edge embeddings $\theta_{\text{edge}}^{(ij)}$, that condition the message passing function $m_{\theta_m}$. 
\begin{remark}
    In the context of power systems, the node embeddings allow for different dynamical properties of nodes. 
    Such properties can be time constants, inertias of a generators, gains, operating regimes or also loads attached to a node. 
    The edge embeddings allow for different line properties, e.g., different conductances and susceptances.  
\end{remark}

\subsection{Initial state encoder $h_{\theta_h}$}
\label{sec:encoder_tasks}

In many system identification problems, the state $x^{(i)}(t_s)$ is not fully measured. 
This is also the case in \eqref{eq:nonlinear_state_model}.
Inspired by \cite{forgione_learning_initial_2022_arx}, a sequence model is used to map past measurements and control inputs to an initial state. 
In this context, the inferred state does not necessarily coincide with the true system state, but is rather a latent representation used to reproduce the observed trajectories with \eqref{eq:message_passing_neural_ode}.

Without loss of generality, we partition the output of each node $y^{(i)}(t) = [\tilde{x}^{(i)}(t)^T ~ \tilde{y}^{(i)}(t)^T]^T$, into measured states $\tilde{x}^{(i)}(t)$ and remaining outputs $\tilde{y}^{(i)}(t)$. 
We then augment the unmeasured states of each node by inferring latent representations of their true initial values with a learned function $\tilde{h}_{\theta_{h}}(\xi)$ at node $i$
\begin{align}
    \big[h_{\theta_h}(\xi)\big]^{(i)} = \begin{bmatrix}
        \pi_x(\xi^{(i)}) \\
        \big[\tilde{h}_{\theta_{h}}(\xi)\big]^{(i)}
    \end{bmatrix},
\end{align}
where  $\pi_x(\xi^{(i)}) = \tilde{x}^{(i)}(t_s)$ is a selector of the states in measured outputs at the initial time. 
For this, we use $H$ historic observations. 
With a slight abuse of notation to match the interface of later used sequence models, we then have a partially latent initial state of the form
\begingroup
\delimiterfactor=1000
\delimitershortfall=0pt
\begin{align}
    \label{eq:augmentation}
    x^{(i)}(t_{s}) &= \begin{bmatrix}
        \tilde{x}^{(i)}(t_s) \\
        \biggl[\tilde{h}_{\theta_h}\biggl(
            \begin{bmatrix}
                u(t_{s-H}) & \cdots & u(t_{s-1}) \\
                y(t_{s-H+1}) & \cdots & y(t_s)
            \end{bmatrix}
        \biggr)\biggr]^{(i)}
    \end{bmatrix}.
\end{align}
\endgroup

\subsection{Neural Structures}
\label{sec:dynamics_models}

For the learned augmentation function $\tilde{h}_{\theta_h}$, we use \acp{stgcn}. 
From the historic observations of all nodes $i \in \mathcal{V}$, we jointly encode a sequence through multiple residual blocks as described in Section \ref{sec:tcn}. 
To the last time instant of the output sequence, we apply a shared node-wise projection head, i.e., a linear layer, to obtain the latent initial state at each node. 

For the node dynamics $f_{\theta_f}$ and the message passing function $m_{\theta_m}$, we use fully connected \acp{mlp}. 
We assume that all measured outputs are also states of the system, i.e., the output function is assumed to be known and is thus just a selector of the respective states, i.e., $y^{(i)}(t) =g_{\theta_g}(x^{(i)}(t)) = \pi_y(x^{(i)}(t))$ of each node. 
\begin{remark}
    If next to the states, the output contains additional variables, the output function can easily be extended with a learnable part, mapping the latent state space of the \ac{node} to these variables, i.e., $y^{(i)}(t) = g_{\theta_g}(x^{(i)}(t)) = [\pi_y(x^{(i)}(t))^T ~ \tilde{g}_{\theta_g}(x^{(i)}(t))^T]^T$.
\end{remark} 

\subsection{Unrolling the model}
\label{sec:unrolled_model}

In control problems, the control input $u(t)$ usually varies over time.
Therefore, the continuous time interval $[t_{s}, t_{e})$ of \eqref{eq:general_solution} can contain discontinuous changes in $u$ over the time horizon.
We therefore employ an autoregressive prediction scheme. 
Consider the sample times $t_{s}, \allowbreak t_{s+1}, \allowbreak \dots, \allowbreak t_{s+K} \in [t_{s}, t_{e})$, where $K$ is the number of prediction steps and the control inputs are constant in the interval $[t_{s}, t_{s+1})$.
Then, we proceed as follows: 
At time $t_{s}$, we jointly encode the initial states of all nodes from $H$ past measurements of $u$ and $y$ as described in Section~\ref{sec:encoder_tasks} and solve the \ac{ivp} to obtain $x^{(i)}(t_{s+1})$. 
For subsequent time intervals $[t_{s+k}, t_{s+k+1}), \, k = 1, \dots, k = K-1$, we use the solution $x^{(i)}(t_{s+k})$ as the initial state for the \acp{ivp} and condition $f_{\theta_f}$ on $u^{(i)}(t_{s+k})$ to find $x^{(i)}(t_{s+k+1})$. 
That way, we recursively solve \acp{ivp} to obtain predictions $\hat{y}^{(i)}(t_{s+1}), \dots, \hat{y}^{(i)}(t_{s+K})$. 
An illustration of the overall unrolled prediction model is shown in Figure~\ref{fig:neural_ode_stepper}.

\subsection{Training the model}
\label{sec:training_procedure} 
We train the model on the dataset $\mathcal{D} = \{(\xi_n, \eta_n)\}_{n=1}^N$ with input features and targets among different time instants $t_{s_n}$. 
The dataset can be partitioned with respect to the nodes into $\xi^{(i)}_{n}$ and $\eta^{(i)}_{n} $ $\forall i \in \mathcal{V}$, i.e., $\xi_n = [\xi^{(1)}_{n} ~\cdots ~ \xi^{(|\mathcal{V}|)}_{n}]$ and $\eta_n = [\eta^{(1)}_{n} ~ \cdots ~\eta^{(|\mathcal{V}|)}_{n}]$. 
These then represent input features and targets of the form 
\begin{align*}
 \xi^{(i)}_n = &[u^{(i)}(t_{s_n-H})^T ~ \cdots ~ u^{(i)}(t_{s_n+K})^T ~ \\ & \,\, y^{(i)}(t_{s_n-H+1})^T ~ \cdots ~ y^{(i)}(t_{s_n})^T]^T, \\
 \eta^{(i)}_n = &[y^{(i)}(t_{s_n+1})^T ~ \cdots  ~y^{(i)}(t_{s_n+K})^T]^T.
\end{align*}
In the training, we aim to minimize the \ac{mse} between the targets $\eta^{(i)}_n$ and the node-wise predictions resulting from $\xi_n$ using the loss function 
\begin{equation}
    \label{eq:mse}
    L = \dfrac{1}{|\mathcal{V}| N K}\sum_{i=1}^{|\mathcal{V}|} \sum_{n=1}^{N} \sum_{k=1}^{K} ||y^{(i)}(t_{s_n+k}) - \hat{y}^{(i)}_n(t_{s_n+k})||_2^2.
\end{equation}
Note that, we use the "discretize-then-optimize" approach \citep{kidger_neural_2022} to obtain $\partial L / \partial \theta$, where $\theta = [\theta_f, \theta_h, \theta_{\text{node}}, \theta_{\text{edge}}]$. 

\subsection{Changing the topology}
As all functions in the \acp{mpgnode} are shared, the topology of the model can be easily modified. 
This allows us to transfer previously learned models to changed topologies. 
Nonetheless, as each node and edge have their own embedding, it is necessary to find the embeddings for an added node or edge. 
This can be done, e.g., by choosing it equivalent to the embedding of a certain node or edge with known or inferred similarity. 
If no similarity measure is available, the embedding can also be initialized based on the mean of all embeddings.
Except for that, the adjacency information in form of each nodes' neighbors $\mathcal{N}_i \, \, \forall i \in \mathcal{V}$ and the normalized adjacency matrix used by the encoder have to be updated. 
Removing nodes or edges requires the same adjacency updates.
When adding or removing nodes, the encoder input is extended or reduced accordingly. 
Note that in all cases accurate predictions are not guaranteed and retraining on data from the altered system may be necessary. 
However, retraining effort is typically much smaller than training an entirely new model from scratch. 


\section{Case Study}
\label{sec:case_study}

\begin{figure}[h]
    \centering
    \vspace{2mm}
    \includegraphics{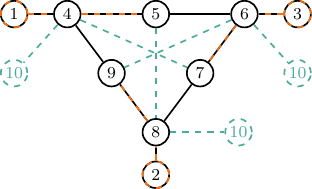}
    \caption{Graph of the IEEE 9-bus system. 
    The unmodified topology is shown in black. 
    Dashed green components indicate additions in the transfer learning experiments; dashed orange overlays indicate removals.}
    \label{fig:graph_ieee9}
\end{figure}

In what follows we use the model \eqref{eq:message_passing_neural_ode} to identify the dynamics of a power system. 
As a test system, we consider an altered version of the IEEE 9-bus system with 9 generators, rendering its dynamics significantly more complex.
The unmodified graph of the system is shown in Figure \ref{fig:graph_ieee9} in black color. 
Each node is equipped with a unit with dynamics \eqref{eq:unit_model}.
The time constants at node 1-3 are $\tau_i\sim\mathcal{U}(0.9,1.1)$ resembling small conventional units, and $\tau_i\sim\mathcal{U}(0.3,0.5)$ at nodes 4-9, resembling fast power electronics-based units. 
The droop gains of all units are $k_i^p = 1$ and $k_i^q = 0.1$.
The network topology, nominal active and reactive power setpoints $p_i^{d,\text{nom}}$ and $q_i^{d,\text{nom}}$ of the units, as well as the network admittances are retrieved from Matpower \citep{zimmermann_matpower_2011}. 

All models are implemented and simulated in Python. 
For the \acp{node}, we use the torchdiffeq package \citep{torchdiffeq} to solve \acp{ivp}, employing a fixed step fourth order Runge-Kutta method. 
As described in Section~\ref{sec:overall_model}, we assume that the voltage and the frequency of each node can be measured while the phase angle measurements are not available. 
We assume zero-mean additive Gaussian measurement noise, see Figure \ref{fig:trajectories}.

\subsection{Setup} 
\label{sec:setup}
\subsubsection{Simulation:}
We generate data by simulating distinct system trajectories of length \SI{1000}{s}. 
As control inputs, we use the active power setpoints of all units. 
For each unit we use random step changes in the active power setpoints $p_i^{d,\text{nom}} + \Delta p_i^d$ where $\Delta p_i^d$ are uniformly distributed around their nominal setpoint, i.e., $\Delta p_i^d \sim  \mathcal{U}(\SI{-0.2}{pu},\SI{0.2}{pu})$. 
These stepwise changes occur every \SI{5}{s}. 
\begin{remark}
    In practice, such step changes are triggered by a higher level controller on the timescale of minutes rather than seconds \citep{hans_operation_2021}. 
    However, we build datasets by cutting out samples of shorter length. 
    Such samples can be realistically obtained by selecting the time windows of interest.  
\end{remark}

\subsubsection{Datasets:}
In order to obtain datasets $\mathcal{D} = \{(\xi_n, \eta_n)\}_{n=1}^N$, we create samples similar to \cite{forgione_continuous_2021}: 
At a sampling rate of $\SI{100}{Hz}$ we slice out $H$ historic observations and $K$ future observations around time instants $t_{s_n}$. 
The number of samples between time instants $t_{s_n}$ and $t_{s_{n+1}}$ is $D$. 
From those slices, we form input features $\xi$ and targets $\eta$ as described in Section~\ref{sec:training_procedure}. 
By doing this for distinct trajectories, we obtain distinct datasets $\mathcal{D}_1-\mathcal{D}_4$. 
We use $\mathcal{D}_1$ for training and $\mathcal{D}_2$ for validation, i.e., early stopping of the training and choosing the best model over all epochs. 
Dataset $\mathcal{D}_3$ is used to choose the best model from 10 different training runs with different parameter initialization seeds, and $\mathcal{D}_4$ as an entirely unseen test set to evaluate the final model. 
For $\mathcal{D}_1-\mathcal{D}_3$ we choose $H = 64$, $K = 256$ and $D = 16$. 
In order to analyze the predictive performance on changes in the control inputs, we choose $H = 64$, $K = 500$ and $D = 500$ for $\mathcal{D}_4$.
Here, $t_{s_n}$ is chosen to be equal to the time instants of the step changes in $u$. 

For each transfer learning experiment, i.e., each topology change, we generate another three datasets $\mathcal{D}_1^{\text{TL}}-\mathcal{D}_3^{\text{TL}}$.
$\mathcal{D}_1^{\text{TL}}$ and $\mathcal{D}_2^{\text{TL}}$ are used for retraining and are constructed similarly to $\mathcal{D}_1$ and $\mathcal{D}_2$ but under differently seeded control inputs. 
Dataset $\mathcal{D}_3^{\text{TL}}$  is again used for evaluation and is constructed similarly to $\mathcal{D}_4$ with trajectories resulting from the same control-input trajectories at nodes retained after topology changes.

\subsubsection{Training:}
For the training, we make use of mini-batch gradient descent with a batch size of 256 using the Adam optimizer \citep{kingma_adam_2017}. 
We choose the following \ac{nn} sizes: 
The graph dynamics $f_{\theta_f}$ as well as the message $m_{\theta_m}$ are made of 2 hidden layers with 128 hidden neurons each where a sigmoid linear unit activation function is used. 
The number of residual blocks of the encoder is set to 4 and its number of hidden channels to 128. 
Furthermore, we consider a learning rate of $0.001$.
The node and edge embeddings' dimension is fixed to 4 and the message dimension is fixed to 8. 

\subsection{Comparison with Baseline Models}
To analyze the capabilities of the \ac{mpgnode} for system identification, we first compare its prediction accuracy to a baseline monolithic \ac{node} of type \eqref{eq:general_neural_ode} which uses a \ac{tcn} encoder. 
A more thorough description of this baseline model can be found in \cite{wolf_augmented_2025}. 
To account for additional complexity of the graph model through two networks $f_{\theta_f}$ and $m_{\theta_m}$, we choose 256 hidden neurons in the dynamics function of the monolithic \ac{node}. 
All remaining hyperparameters are chosen equivalent to the \ac{mpgnode}.
Moreover, to evaluate the effectiveness of embeddings that enable heterogeneity, we compare the \ac{mpgnode} to a homogeneous \ac{mpgnode} which does not make use of such embeddings.

\begin{figure}[h]
    \centering
    \includegraphics{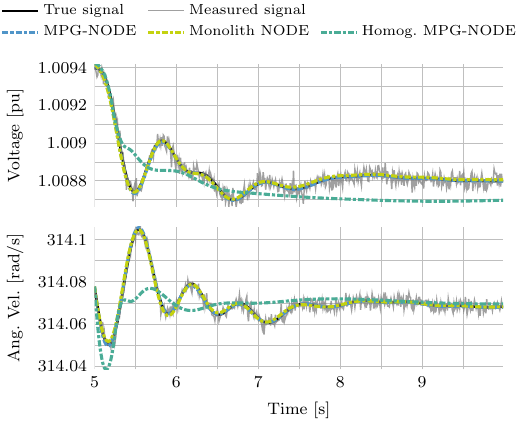}
    \caption{Example predictions of the IEEE 9-bus system on node 6 for the \ac{mpgnode} and the baseline models.
    }
    \label{fig:trajectories}
\end{figure}

\begin{figure}[h]
    \centering
    \vspace{1.5mm}
    \includegraphics{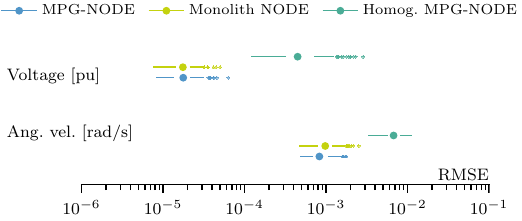}
    \caption{Boxplot of voltage and frequency \acp{rmse} of the evaluation dataset of the IEEE 9-bus system for the \ac{mpgnode} and baseline models. 
    }
    \label{fig:boxplot_comparison}
\end{figure}

In order to compare the three models, we analyze the \ac{rmse} of the prediction on the evaluation dataset $\mathcal{D}_4$. 
With the chosen $K$, we obtain a prediction of the voltages and frequencies over \SI{5}{s} after a step change in the control input. 
Figure~\ref{fig:trajectories} shows example predictions of the system trajectories of the \ac{mpgnode}, monolithic \ac{node}, and homogeneous graph \ac{node} for a single sample of the evaluation dataset.
Figure~\ref{fig:boxplot_comparison} shows the distribution of the \ac{rmse}, i.e., the root of \eqref{eq:mse}, separated into voltages and frequencies. 
For both, the \ac{mpgnode} and the monolithic \ac{node} yield similar performance. 
Both can accurately reproduce the true trajectories of the system, with only slight dynamic deviations and steady-state errors. 
Leveraging the graph structure as an inductive bias did not yield significant improvements in our experiments. 
That being said, an already very accurate baseline model was chosen and thus the room for improvement is small. 
For both models, all seeds resulted in successful training, i.e., accurate fits with comparable performance. 
However, the homogeneous graph \ac{node} could not identify the heterogeneous dynamics of the system and resulted in significantly larger errors.


\subsection{Transfer learning}

\begin{figure}[h]
    \centering
    \includegraphics{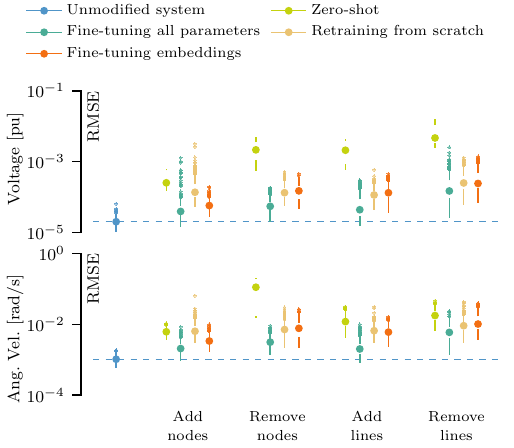}
    \caption{Voltage and frequency \acp{rmse} on transfer learning experiments. Each topology-change category pools three experiments. The unmodified-system reference is shown once on the left. 
    }
    \label{fig:boxplottransferpooled}
\end{figure}

The \ac{mpgnode} offers a much more flexible structure than the monolithic \ac{node} as nodes and edges can be added or removed without relearning an entire model. 
Opposed to that, the structure of the monolithic \ac{node} is fixed, and requires rebuilding and retraining the model after topology changes. 
In what follows, we illustrate these transfer learning capabilities on four kinds of experiments: Adding one new node (to node  $4$, $6$ and $8$) and a respective edge, removing one node ($1$, $2$ and $3$) and its associated edges, as well as adding one edge ($(4,7)$, $(5,8)$ and $(6,9)$) and removing one edge ($(4,5)$, $(6,7)$ and $(8,9)$) between existing nodes. 
Figure~\ref{fig:graph_ieee9} highlights the topology changes.
When adding a new node to the system, we associate it with a fast power electronics-based unit with $\tau_{10} \sim \mathcal{U}(0.3,0.5)$. 
When adding an edge, we associate it with the line parameters $b=-10$ and $g=1$. 

We create new trajectories and datasets as described in Section \ref{sec:setup}. 
We assume no knowledge of similarity and initialize the edge and node embeddings for the unseen components by averaging across all embeddings learned on the unmodified system. 
We analyze the retraining results of the \ac{mpgnode} without (zero-shot) and with retraining the model on a 10\% fraction of the original dataset. 
As a reference, we use the prediction accuracy of the unmodified system. 
Three kinds of retrainings are analyzed:
1) Fine-tuning all parameters of the model.
2) Retraining the model from scratch.
3) Fine-tuning only the embeddings with frozen neural networks $f_{\theta_f}$ and $m_{\theta_m}$.
The evaluation is done on the full dataset $\mathcal{D}_3^{\text{TL}}$. 
Figure~\ref{fig:boxplottransferpooled} shows the distribution of the \ac{rmse} of the voltage and frequency predictions for all four categories of transfer learning experiments for zero-shot and all retraining approaches together with the reference.

\subsubsection{Fine-tuning all parameters:}
While adding nodes resulted in relatively accurate zero-shot predictions, removing nodes or edges resulted in a significant deterioration of the prediction accuracy. 
This is due to experiment design: the added nodes carry no additional load and operate as storage units with power setpoints distributed around a nominal setpoint of zero. 
They therefore do not significantly change the power flow through the network but rather induce local changes in operating regimes of adjacent nodes. 
On the other hand the removed nodes are major power suppliers in the system. 
Removing them therefore entirely changed the power sharing and operating regimes of the remaining nodes. 
The same applies to the removal and addition of edges on important paths of the network which changed the power flow and therefore the power injections at many nodes of the system. 
However, even with only 10\% of the original dataset size, retraining was able to significantly reduce the mismatches resulting from the topology changes such that the accuracy got close to the unmodified system. 

\subsubsection{Retraining from scratch:}
Comparing the retraining from scratch with the fine-tuning of all parameters, it shows that the latter produces more accurate results in all experiment categories. 
This clearly indicates that learned knowledge is successfully transferred to the changed topology. 

\subsubsection{Fine-tuning embeddings:}
While the fine-tuning of the full model could successfully adapt to the topology changes, fine-tuning only the embedding does not yield a comparable performance. 
While a meaningful representation of the shared dynamics is learned and there is a significant improvement compared to the zero-shot predictions, the embeddings alone appear insufficient to adapt to the changed model dynamics. 
However, it still performs slightly better or similarly well than training from scratch across all experiments.
Note that convergence was significantly slower than for all other retraining approaches, such that the learning rate was chosen 30 times larger.


\section{Conclusion}
\label{sec:conclusion}

In this paper, we developed a novel \ac{mpgnode} for heterogeneous node dynamics and edge couplings to identify the coupled dynamics of power systems. 
The model was trained on data from the IEEE 9-bus system and achieved similar prediction accuracy as the baseline monolithic \ac{node}. 
Furthermore, it was shown that the embeddings allowed to identify the heterogeneous dynamics on the graph where a comparable homogeneous \ac{mpgnode} failed.
In addition, it was shown to offer a much more flexible structure that allows for easy adaptation to changes in the system topology: the \ac{mpgnode} could be successfully adapted to a system with an added node and edge with only a fraction of the original training data and effort. 
In future work we want to investigate potential improvements in the prediction accuracy by employing physics-inspired regularization in the edge couplings. 
Furthermore, we want to investigate improved zero-shot transfer learning potential by combining persistent and sample-specific encoded embeddings.  

\bibliography{literature}

\end{document}